\documentclass{aip-cp}
\usepackage[numbers]{natbib}
\usepackage{rotating}
\usepackage{graphicx}

\usepackage[caption=false]{subfig}

\newenvironment{subfigures}
 {\begin{minipage}{\columnwidth}\def\@captype{figure}\centering}
 {\end{minipage}}
\makeatother

\begin{document}

\title{Studies Of The Polarization Of Gamma Photons Originating From The Decay of Positronium Atoms}

\author{Sushil K. Sharma\corref{cor1}}
\author{Nikodem Krawczyk}
\author{Juhi Raj for the J-PET collaboration}

\affil[aff1]{The Marian Smoluchowski Institute of Physics, Jagiellonian University, Lojasiewicza 11, 30-348 Krakow, Poland}
\corresp[cor1]{Corresponding author: sushil.sharma@uj.edu.pl}

\maketitle

\begin{abstract}
The precise measurements of the Compton scatterings of photons originating from the decay of positronium atoms can reveal information about their polarizations. J-PET detector is constructed of 192 plastic scintillators and is unique to study the scattering correlations of the annihilation photons with an angular precision of several degrees. In this work, we present the first experimental evidence showing the feasibility of measuring the photons relative polarization using the J-PET detector.     
\end{abstract}
\section{INTRODUCTION}
Positronium is a bound state of electron and positron which resembles the hydrogen atom but without the presence of a nucleus. Photons emitted in the decay of positronium atom possesses the orthogonal linear polarization states due to the zero relative angular momentum of the electron-positron pair just prior to the decay~\cite{DIR30,HAR47}. There is no direct way to measure the polarization of these high energetic photons. However, the studies on the scattering angles as well as azimuth variations in their Compton scatterings can help to evaluate their linear polarization~\cite{SIN65}.
The primary scattering of the incident photon by free electron is most likely at the right angle to its polarization vector~\cite{KLE29,EVA58}. Based on that, one can define the direction of the polarization vector~\cite{MOS16B} as $\vec{\epsilon} = \vec{k} \times \vec{k'}$, where $\vec{k}$ and $\vec{k'}$ are momentum vectors of photon before and after the scattering. Studying the angular correlation between the scattering planes of photons on event-wise basis gives access to measure the relative polarization of the photons~\cite{HIE17A}.
Information on the polarization degree of freedom of photons resulting from the decay of positronium atom allow to study the various exotic physical aspects such as  multi-particle entanglement~\cite{HIE17B,NOW17A}. Moreover, this information may also be used to investigate discrete symmetries~\cite{MOS16B,ACI01A} in the decay of positronium atom. In this work, we present the road map to perform the studies on the polarization determination of annihilation photons via Compton scattering on the event-wise basis.
\section{The J-PET detector}
The Jagiellonian Positron Emission Tomograph(J-PET) composed of three layers of 192 plastic scintillators arranged co-axially at the angular displacements such that the scintillators are not shadowing each other from the central view of detector. Each scintillator has a volume of 500 x 19 x 7 mm$^3$ and is made of EJ-230 material connected with R9800 Hamamatsu photomultiplier(PMT) at each end. The scintillators are placed in consecutive layers at radii 42.5, 46.75 and 57.5 cm. The first and second layers have 48 modules whereas the third layer has 96 scintillating modules. More information on technical details about the J-PET can be found in the works~\cite{MOS14A,MOS15A,MOS16A,NIE17A,KOW18A}.
Hit-position and the hit-time of the photon's interaction within the scintillator is estimated based on the time difference in the arrival of light signals to the attached PMTs~\cite{MOS15A,SHA15A,RAC14A,RAC17A}. For the 192 scintillator modules, there are total 384 analog signals to be processed through a trigger-less data acquisition which is based on the time information of the signals instead of energy~\cite{KOR16A,KOR18A}. For the presented study, a point like source $^{22}$Na with 1MBq activity was used.
\begin{figure}[!htp]
    \begin{subfigures}
    {
        \includegraphics[width=8.3cm,height=7.cm,keepaspectratio]{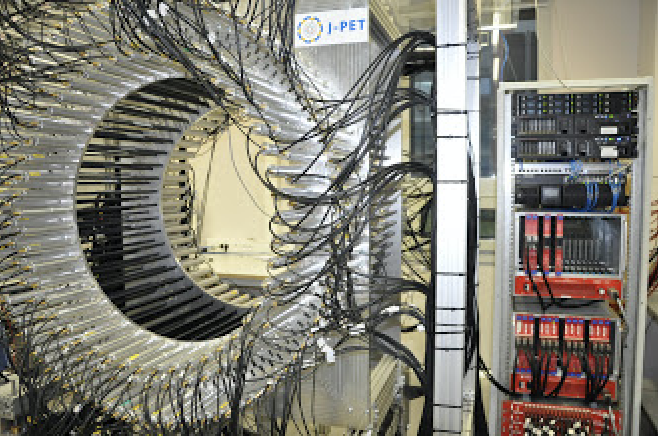}
         \hspace{0.3cm}
        \includegraphics[width=7cm,height=5.5cm,keepaspectratio]{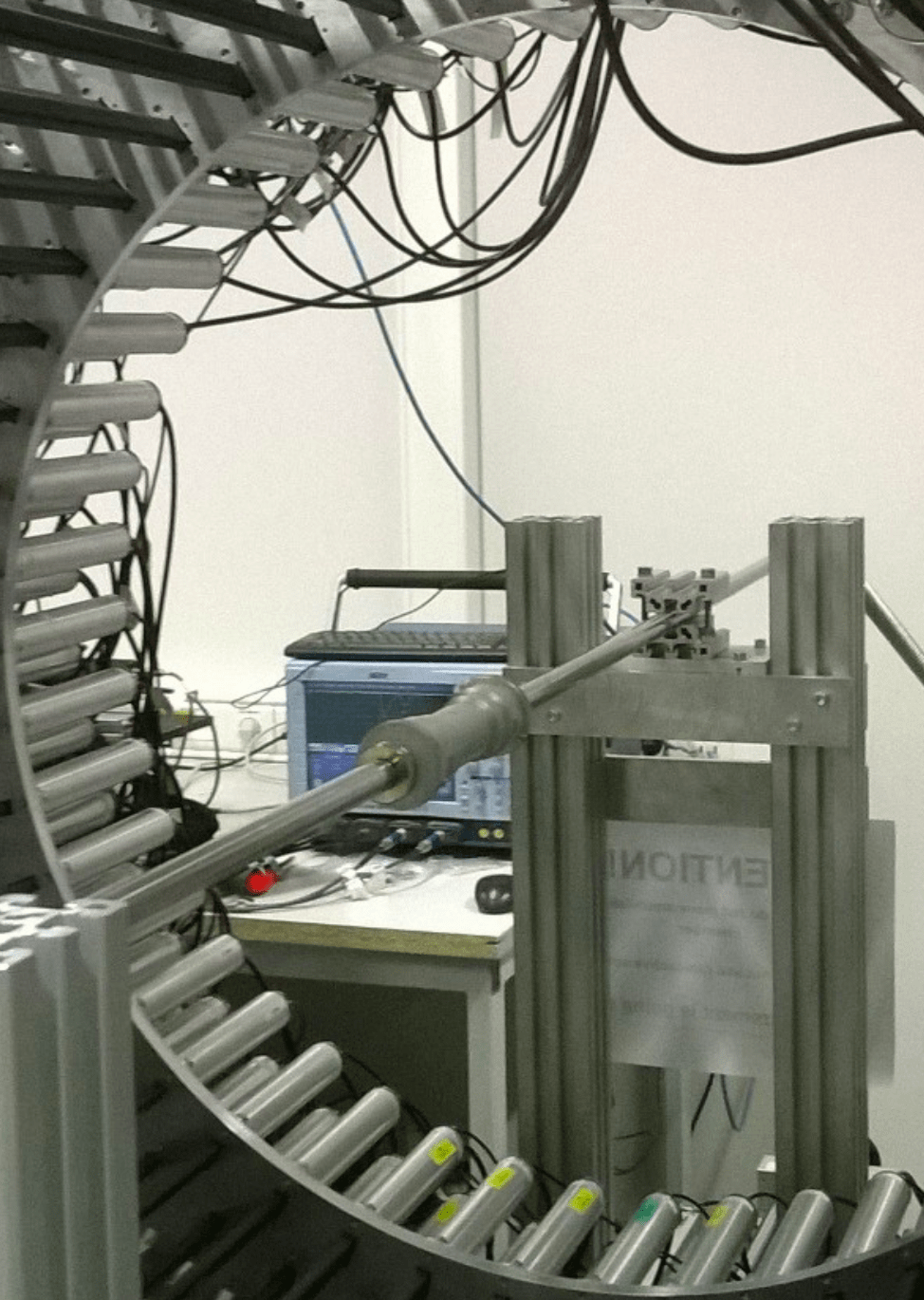}
    } 
    \end{subfigures}
    \caption{Left picture shows the J-PET detector prototype with 3 layers and the right picture presents the placement of the small annihilation chamber in the center of the detector.}
\end{figure}
The source was sandwiched in XAD4 porous material and enclosed in a small chamber of the diameter of $\approx$~5 cm. A controlled monitoring of temperature, pressure and humidity in laboratory was ensured with meteo station. The data was analysed off-line by means of a dedicated framework developed and optimized locally by the J-PET collaboration~\cite{KRZ15A,KRZ15B}. Time-Over-Threshold (TOT) as the estimation of energy deposition by the incident photon and its application in J-PET is discussed separately in the next section.
\section{Relationship between TOT and Energy Deposition}
TOT technique is widely being adopted for the multi-channel readouts particularly dealing with the fast timing signals~\cite{CAR15A,GRA14A}. The main idea is that one basically measures the pulse width which is somewhat the estimate of the signal's amplitude (i.e., charge collection). Thus the TOT approach transforms the voltage domain to the time measurements. This approach despite having advantages as the compactness of signals readout and power consumption confronts a challenge in terms of non-linear input energy to pulse width conversion. In the framework of J-PET, charge collection was replaced by TOT which is the measure of an amount of light registered by the PMTs at both ends. Each analog signal is probed at four thresholds for dealing with the nonlinearity and improving the energy resolution~\cite{NIE17A,PAL17A}. To use the potential of the TOT technique, an algorithm was developed to establish the relationship between the measured TOT and energy deposition by the incident photon. 
\begin{figure}[!htp]
    \begin{subfigures}
    {        \includegraphics[width=10cm,height=4.5cm,keepaspectratio]{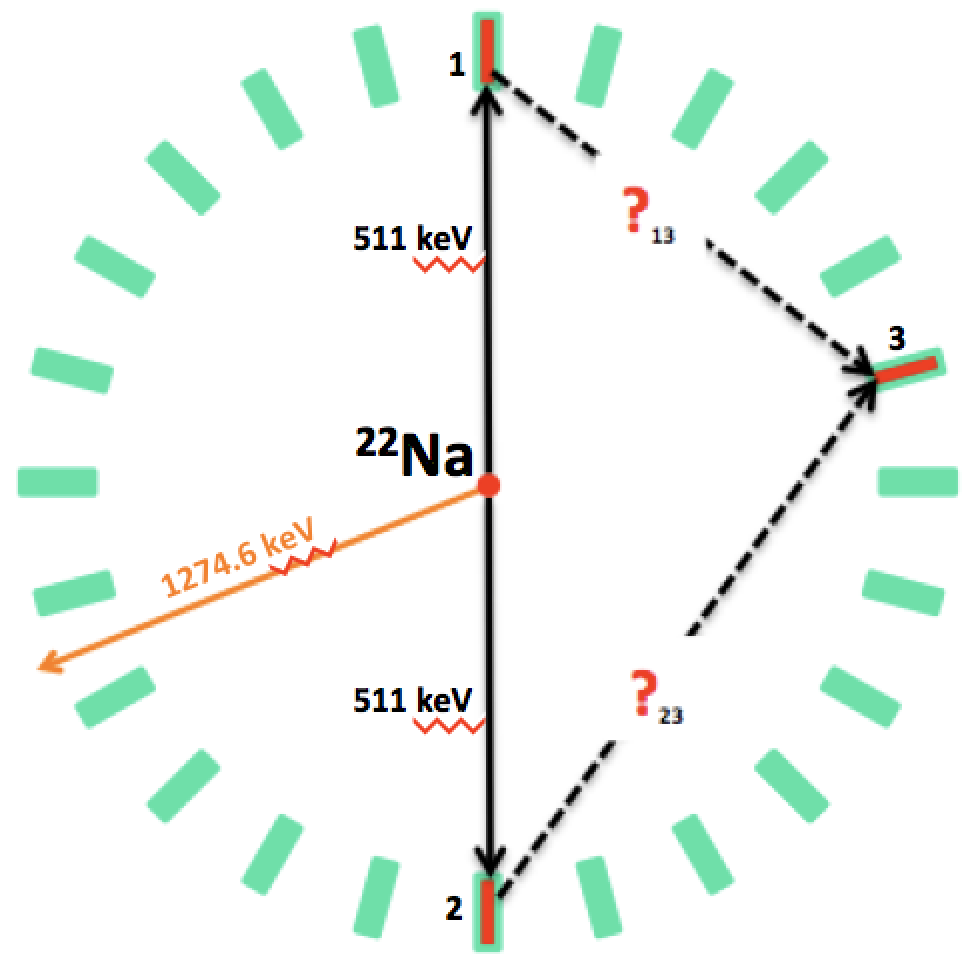}
     \hspace{1.5cm}
        \includegraphics[width=8cm,height=4.5cm,keepaspectratio]{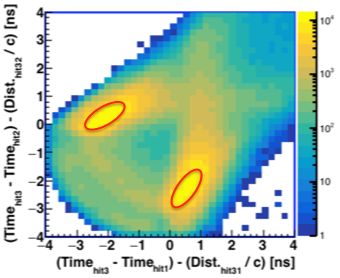}
    } 
    \end{subfigures}
    \caption{Selection criterion for the 511 keV photon is shown in left picture. The interaction of back-to-back photon is shown by 1 and 2 whereas the 3 is showing the interaction of scattered photon with scintillator. Right picture shows the scatter test~(S) results for the assignment of scattered photon to its primary interaction.}
    \label{511keV}
\end{figure}
Such a relationship may help to ensure the efficient identification of particle originating from the different sources (e.g., from $^{22}$Na deexcitation, positronium decay etc.) which is essential for the studies of discrete symmetries and photon's polarization. The method adopted to attain this relationship is explained based on the identification of 511 keV photon originating from the e$^+$- e$^-$ annihilation. As the plastics are made of hydrocarbon materials, high energetic photons interact predominantly via the Compton effect. For this study, 3 hits (hit refers to an interaction of photon with scintillator) events were studied where the first two hits are due to the 511 keV photons. The selection of 511 keV photons can be easily ensured based on the back-to-back criterium due to the momentum-energy conservation. The schematic of the 3-hits event is shown in left panel of Figure~\ref{511keV}. The third hit represents the interaction of the scattered one which can be the outcome of either of 511 keV photons.  
The assignment of the scattered photon to its primary photon is crucial as it allows to estimate the scattering angle. Knowing the precise scattering angle and the incident energy of the primary photon, one can calculate the energy deposition in interaction within the scintillator~\cite{COM23A}. The assignment of the scattered hit was done based on the results of the scatter test~(\textbf{S}) which measure the deviation between the measured hit-time and time calculated using the measured distance between the hit-positions divided by the speed of light (right panel of Figure~\ref{511keV}). For the ideal case, the value of this test should be zero. But considering the hit-time and hit-position resolutions the value of S-test can be smeared. So one can select the time window for the S-test assuming the selection of true scatterings (an area enclosed by red line in Figure~\ref{511keV}).  
Thus, for any primary hit, we can measure the TOT value and estimates the corresponding energy deposition based on the incident energy of photon and its scattering angle to obtain the relationship between TOT and energy deposition. In order to extend such relationship for higher energies, the scattering studies of the 1274.6 keV photons which are also emitted from the $^{22}$Na source were performed. The detailed analysis procedure and the outcomes from TOT studies are the subject of the future work, here we chose to present only the methodology showing how the J-PET detector can be calibrated in terms of energy based on the photon's scattering. In the next section, the idea to study the polarization of annihilation photons by means of the J-PET detector is described. 
\section{Measurement of photons polarization}
\begin{figure}[!htp]
    \begin{subfigures}
    {        \includegraphics[width=6cm,height=6.5cm,keepaspectratio]{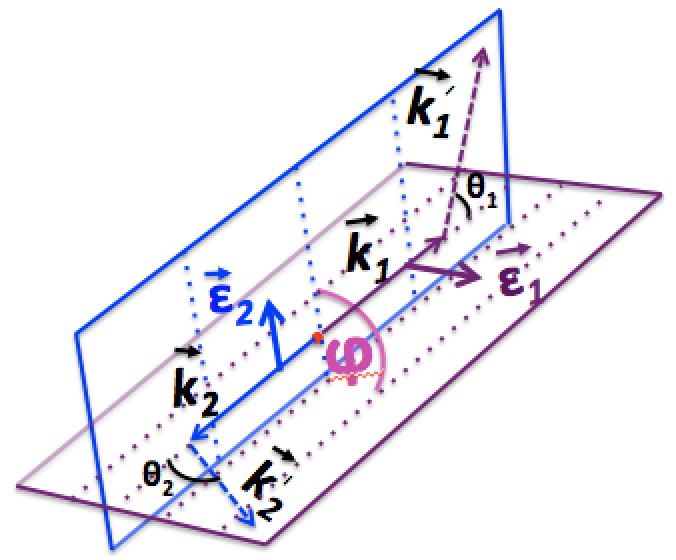}
    \hspace{1cm}
        \includegraphics[width=6cm,height=6.5cm,keepaspectratio]{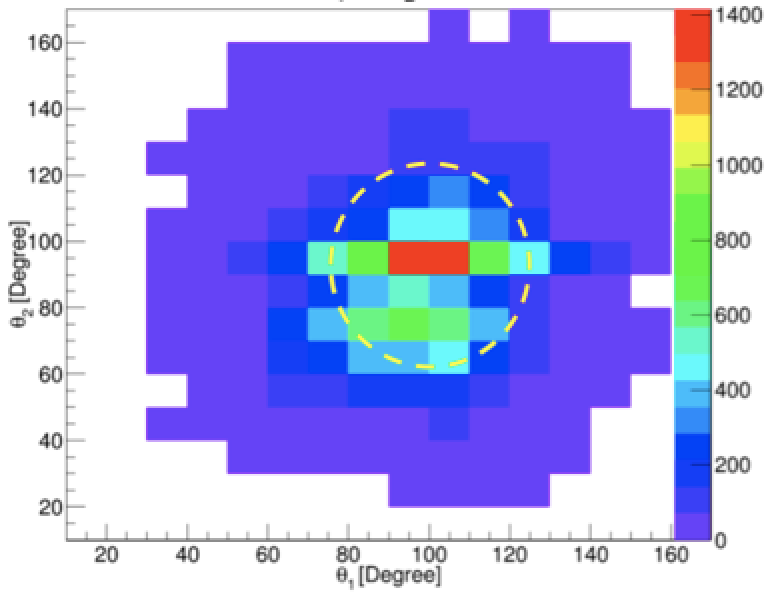}
    } 
    \end{subfigures}
    \caption{Left panel shows the schematic to study the relative polarisation of annihilation photons. Right panel presents distribution of scattering angles $\theta_1$ vs $\theta_2$. It shows that the acceptance of the J-PET detector is maximal at about $\theta_1$ = $\theta_2$ $\approx$ 100$^\circ$ which is close to the $\theta_1$ = $\theta_2$ = 82$^\circ$ where the highest visibility for the polarisation determination is expected~\cite{PRY47A}.}
    \label{polarization}
\end{figure}

With the ability to measure the scattering angles of interacting photons within the precision of several degrees, J-PET proves the capability to perform the tests on measuring the relative polarization of the annihilation photons. Before presenting the first experimental evidence showing the potentialities of the J-PET detector for such studies, the theoretical background is explained in the left panel of the Figure~\ref{polarization}. It is shown that the annihilation photons are emitted in opposite direction in order to preserve the momentum conservation and scattered at angles~$\theta_1$ and~$\theta_2$ respectively. From theoretical calculatoions we know that the polarization vectors of the annihilation photons are perpendicular to each other~\cite{DIR30,HAR47}. This relationship between their polarization cannot be measured directly. However, it can be studied by measuring the scattering angles of the annihilation photons. Photons have maximum probability to be scattered in the perpendicular direction to their polarization vector ( $\vec{\epsilon}$ ). Therefore, a measurable correlation between the scattering distribution of annihilation photons should be visible, which can be considered as an angle between the scattering planes formed by the momentum vector of photon before ($\vec{k}$) and after ($\vec{k'}$) the scattering as shown by $\varphi$ in Figure~\ref{polarization} (left panel). 
For such studies events with only 4 hits were analyzed, where the first two hits were by the back-to-back photons coming from the e$^+$- e$^-$ annihilation and the other two are the hits by the respective scattered photons. The \textbf{S} test was employed for the assignment of the scattered photons. Scattering angles were measured for both photons on event-wise basis in order to study the correlation among them. Figure~\ref{polarization} (right panel) shows the preliminary experimental results. The results are cumulative of all the possible processes responsible to produce the photons e.g., direct annihilation, the decay of para-positronium atom, pick-off reaction. However, using $^{22}Na$ source (with the measurement of the 1274.6 keV deexcitation photon) it will be possible to study the correlation between scattering angles for each of the mentioned processes independently. It will be of great interest to study such angular correlation in the scattering angles for the independent sources of annihilation photons. 
\section{Summary and outlook}
J-PET detector is based on the plastic scintillators optimized to study the scatterings of the high energetic photons, which allows to study the polarization effects in the Compton scattering of annihilation photons resulting from the decay of positronium atom. In plastic, the photon interacts most likely via Compton interaction. The relationship between TOT and the energy deposition by incident photon will allow to estimate the energy deposition for each hit. We have shown that the study to measure the angular correlation of the scatterings of annihilation photons can be performed using the J-PET detector.   
\section{ACKNOWLEDGMENTS}
The authors on behalf of the J-PET collaboration would like to acknowledge the support by The Polish National Center for Research and Development through grant INNOTECH-K1/IN1/64/159174/NCBR/12, the Foundation for Polish Science through the MPD and TEAM/2017-4/39 programmes, the National Science Centre of Poland through grants no. 2016/21/B/ST2/01222, 2017/25/N/NZ1/00861, the Ministry for Science and Higher Education through grants no. 6673/IA/SP/2016, 7150/E- 338/SPUB/2017/1, 7150/E-338/M/2017, 7150/E-338/M/2018 and the Austrian Science Fund FWF-P26783.


\nocite{*}
\bibliographystyle{aipnum-cp}%

\end{document}